\documentclass[12pt,preprint]{aastex}

\usepackage{txfonts}
\usepackage{graphicx}

\begin{document}


   \title{First magnetic field models for recently discovered magnetic $\beta$\,Cephei and slowly pulsating B stars\thanks
{Based on observations obtained at the European Southern Observatory (ESO programme 084.D-0230(A)).}}


\author{S. Hubrig, I. Ilyin}
\affil{Astrophysikalisches Institut Potsdam, An der Sternwarte 16, 14482 Potsdam, Germany}
\email{shubrig@aip.de}

\author{M. Sch\"oller}
\affil{European Southern Observatory, Karl-Schwarzschild-Str.\ 2, 85748 Garching bei M\"unchen, Germany}

\author{M. Briquet}
\affil{Instituut voor Sterrenkunde, Katholieke Universiteit Leuven, Celestijnenlaan 200 D, 3001~Leuven, Belgium}

\author{T. Morel}
\affil{Institut d'Astrophysique et de G\'eophysique, Universit\'e de Li\`ege, All\'ee du 6 Ao\^ut, B\^at. B5c, 4000~Li\`ege, Belgium}

\author{P. De Cat}
\affil{Koninklijke Sterrenwacht van Belgi\"e, Ringlaan 3, 1180 Brussel, Belgium}

%\author{S. Djorgovski\altaffilmark{1,2,3} and Ivan R. King\altaffilmark{1}}
%\affil{Astronomy Department, University of California,
    %Berkeley, CA 94720}
%
%\author{C. D. Biemesderfer\altaffilmark{4,5}}
%\affil{National Optical Astronomy Observatories, Tucson, AZ 85719}
%\email{aastex-help@aas.org}
%
%\and
%
%\author{R. J. Hanisch\altaffilmark{5}}
%\affil{Space Telescope Science Institute, Baltimore, MD 21218}


%\altaffiltext{1}{Visiting Astronomer, Cerro Tololo Inter-American Observatory.
%CTIO is operated by AURA, Inc.\ under contract to the National Science
%\altaffiltext{2}{Society of Fellows, Harvard University.}
%\altaffiltext{3}{present address: Center for Astrophysics,
    %60 Garden Street, Cambridge, MA 02138}
%\altaffiltext{4}{Visiting Programmer, Space Telescope Science Institute}
%\altaffiltext{5}{Patron, Alonso's Bar and Grill}


\begin{abstract}
In spite of recent detections of magnetic fields in a number of $\beta$\,Cephei
and slowly pulsating B (SPB) stars, their impact on stellar rotation, pulsations,
and element diffusion is not sufficiently studied yet.
The reason for this is the lack of knowledge of rotation periods, the magnetic
field strength distribution and temporal variability, and the field geometry.
New longitudinal field measurements of four $\beta$\,Cephei and 
candidate $\beta$\,Cephei stars, and two SPB stars were acquired with FORS\,2 at the VLT.
These measurements allowed us to carry out a search for rotation periods and to
constrain the magnetic field geometry for four stars in our sample.
\end{abstract}


%\keywords{globular clusters: general --- globular clusters: individual(NGC 6397,
%NGC 6624, NGC 7078, Terzan 8}

%\keywords{stars: pre-main sequence --- stars: atmospheres --- stars: individual (HD\,101412) --- magnetic fields }




\section{Introduction}

\begin{table}
\centering
\caption{
The observed $\beta$\,Cephei and SPB stars.
%In the three columns we list the HD number, another 
%identifier, the spectral type retrieved from the SIMBAD database, pulsating type, and membership in a spectroscopic binary system.
%An asterisk in front of the HD number denotes candidate $\beta$\,Cephei stars.
}
\label{tab:table1a}
\begin{tabular}{rccc}
\hline
\hline
\multicolumn{1}{c}{HD} &
\multicolumn{1}{c}{Other} &
\multicolumn{1}{c}{Spectral} &
\multicolumn{1}{c}{Comments} \\
\multicolumn{1}{c}{} &
\multicolumn{1}{c}{Identifier} &
\multicolumn{1}{c}{Type} &
\multicolumn{1}{c}{} \\
\hline
       24587   & 33\,Eri         & B5V     & SPB, SB1\\
       46328   & $\xi^1$\,CMa    & B1III   & $\beta$\,Cep  \\ 
       50707   & 15\,CMa         & B1Ib    & $\beta$\,Cep  \\  
$\ast$ 74575   & $\alpha$\,Pyx   & B1.5III & $\beta$\,Cep \\ 
       74560   & HY\,Vel         & B3IV    & SPB, SB1\\ 
$\ast$ 136504  & $\epsilon$\,Lup & B2IV-V  & $\beta$\,Cep, SB  \\ 
\hline
\end{tabular}
\end{table}

For several years, a magnetic field survey of main-sequence pulsating B-type stars, namely the slowly 
pulsating B (SPB) stars and $\beta$\,Cephei stars, has been undertaken by our team with FORS\,1 in 
spectropolarimetric mode at the VLT, allowing us to detect in four $\beta$ Cephei stars and 
in 16 slowly pulsating B stars, for the first time, 
longitudinal magnetic fields of the order of a few hundred Gauss \citep{Hubrig2006,Hubrig2009}. 
$\beta$~Cephei variables have spectral types B0--B2 and pulsate in low-order
pressure and gravity modes with periods between 2 and 6\,hours.
Slowly pulsating B (SPB) stars are mid-B type (B3--B9) objects pulsating
in high-order gravity modes with periods in the range of 0.5--3\,days. 
Pulsating stars are currently considered as promising targets for asteroseismic analysis
(e.g.\ \citealt{ShibahashiAerts2000}),
which requires as input the observed parameters of the magnetic field topology.
Early magnetic field searches of $\beta$\,Cephei stars were mostly unsuccessful due to low precision
(see \citealt{Babcock1958,RudyKemp1978}).
Before we started our systematic search for magnetic fields in pulsating B-type stars,
a weak magnetic field was detected in two $\beta$\,Cephei stars, in the prototype of the class, $\beta$\,Cep itself, 
by \citet{Henrichs2000} and in V2052\,Oph by \citet{Neiner2003a}.
The first detection of a weak magnetic field in the SPB star $\zeta$\,Cas
was reported by \citet{Neiner2003b}.
The detected magnetic objects 
for which we gathered several magnetic field measurements showed a field that varies in time, but no 
periodicity could be derived yet due to the limited amount of VLT observing time.
Among these targets with a detected magnetic field, we  
selected two slowly pulsating B stars, two $\beta$ Cephei stars, and two candidate $\beta$ Cephei stars with
suitable coordinates, for successive VLT multi-epoch magnetic measurements.
%from 2009 September to 2010 March.
The list of the selected targets is presented in Table~\ref{tab:table1a}. 
In the four columns we list the HD number, another identifier,
the spectral type retrieved from the SIMBAD database, the pulsating type,
and membership in a spectroscopic binary system.
An asterisk in front of the HD number denotes candidate $\beta$\,Cephei stars (cf.\ \citealt{StankovHandler2005}).
This most recent study aimed at the determination of magnetic field properties for these stars, such as 
field strength, field geometry, and time variability.
Here we present the results of 62 new magnetic field measurements of the six selected stars
and discuss the obtained results on their rotation periods and magnetic field geometry.

\section{Magnetic field measurements and period determination}

Multi-epoch time series of polarimetric spectra of the pulsating stars were obtained
with FORS\,2\footnote{The spectropolarimetric capabilities of FORS\,1 were moved to FORS\,2 in 2009.}
on Antu (UT1) from 2009 September to 2010 March in service mode.
Using a slit width of 0\farcs4, the achieved spectral resolving power 
of FORS\,2 obtained with the GRISM 600B was about 2000.
%The $\beta$\,Cephei stars $\xi^1$\,CMa and 15\,CMa were observed 11 and 13 times, respectively.
%For the candidate $\beta$\,Cephei stars $\alpha$\,Pyx and $\epsilon$\,Lup, only a handful of observations
%were obtained. The SPB stars 33\,Eri and HY\,Vel were observed 13 and 12 times, respectively.
 A detailed description of the assessment of the longitudinal 
magnetic field measurements using FORS\,2 is presented in our previous papers 
(e.g., \citealt{Hubrig2004a,Hubrig2004b}, and references therein). 
%We repeat here the major steps of the magnetic field determination. 
The mean longitudinal 
magnetic field, $\left< B_{\rm z}\right>$, was derived using 

\begin{equation} 
\frac{V}{I} = -\frac{g_{\rm eff} e \lambda^2}{4\pi{}m_ec^2}\ \frac{1}{I}\ 
\frac{{\rm d}I}{{\rm d}\lambda} \left<B_{\rm z}\right>, 
\label{eqn:one} 
\end{equation} 

\noindent 
where $V$ is the Stokes parameter which measures the circular polarisation, $I$ 
is the intensity in the unpolarised spectrum, $g_{\rm eff}$ is the effective 
Land\'e factor, $e$ is the electron charge, $\lambda$ is the wavelength, $m_e$ the 
electron mass, $c$ the speed of light, ${{\rm d}I/{\rm d}\lambda}$ is the 
derivative of Stokes $I$, and $\left<B_{\rm z}\right>$ is the mean longitudinal magnetic 
field.  The measurements of the longitudinal magnetic 
field were carried out in two ways,  using the whole spectrum  ($\left<B_{\rm z}\right>_{\rm all}$) 
and using only the hydrogen lines ($\left<B_{\rm z}\right>_{\rm hyd}$).
%using the whole spectrum and the longitudinal magnetic field 
%$\left<B_{\rm z}\right>_{\rm hyd}$ using only the hydrogen lines. 

%To minimise the cross-talk effect, we executed a sequence of spectropolarimetric observations
%at different position angles of the retarder waveplate, +45$-$45, 
%$-$45+45, +45$-$45, etc., and calculated the values $V/I$ using: 
 
%\begin{equation} 
%\frac{V}{I} = 
%\frac{1}{2} \left\{ \left( \frac{f^{\rm o} - f^{\rm e}}{f^{\rm o} + f^{\rm e}} 
%\right)_{\alpha=-45^{\circ}} 
%- \left( \frac{f^{\rm o} - f^{\rm e}}{f^{\rm o} + f^{\rm e}} 
%\right)_{\alpha=+45^{\circ}} \right\}, 
%\label{eqn:two}   
%\end{equation} 
 
%\noindent 
%where $\alpha$ denotes the position angle of the retarder waveplate and $f^{\rm 
%o}$ and $f^{\rm e}$ are ordinary and extraordinary beams, respectively. Stokes $I$ 
%values were obtained from the sum of the  ordinary and extraordinary beams. 
Two additional polarimetric spectra of $\xi^1$\,CMa were obtained with the SOFIN spectrograph 
installed at the 2.56\,m Nordic Optical Telescope on La Palma, one on 2008 September 13,
and another one on 2010 January 01.
SOFIN \citep{Tuominen1999} is a high-resolution echelle spectrograph 
mounted at the Cassegrain focus of the NOT.
%and equipped with three optical cameras providing 
%different resolving powers of 30\,000, 80\,000, and 160\,000.
The star was observed with the low-resolution camera with $R=\lambda/\Delta\lambda\approx$\,30\,000.
We used the 2K Loral CCD detector to register 40 echelle orders partially
covering the range from 3500 to 10\,000\,\AA{} 
with a length of the spectral orders of about 140\,\AA{}.
% at 5500\,\AA{}.
%The polarimeter is located in front of the entrance slit of the spectrograph and consists of a fixed 
%calcite beam splitter aligned along the slit and a rotating super-achromatic quarter-wave plate. Two spectra 
%circularly polarized in opposite sense are recorded simultaneously for each echelle order, providing 
%sufficient separation 
%by the cross-dispersion prism below 7000\,\AA{}. 
Two such exposures with quarter-wave plate angles separated 
by $90^\circ$ are necessary to derive circularly polarised spectra.
The spectra were reduced with the 4A software package \citep{Ilyin2000}. 
%Bias subtraction, master flat-field correction, 
%scattered light subtraction, and weighted extraction of spectral orders comprise the standard steps of the data 
%processing. A ThAr spectral lamp is used for wavelength calibration, taken before and after each target exposure 
%to minimize temporal variations in the spectrograph. 
%%We measured a mean longitudinal magnetic field 
%%$\langle$$B_z$$\rangle$\,=\,386$\pm$39\,G in 2008 September and $\langle$$B_z$$\rangle$\,=\,297$\pm$26\,G
%%in 2010 January. 

%\begin{table}\scriptsize
\begin{table}\tiny
\centering
\caption{
%Magnetic field measurements of $\beta$\,Cephei and SPB stars with FORS\,2.
%Phases are calculated according to the ephemeris of 
%Stahl et al.\ (\cite{stahl:1996}), ${\rm JD} = 2\,448\,833.0 + 15.422~{\rm E}$.
Magnetic field measurements of $\beta$\,Cephei stars with FORS\,1/2 and SOFIN
(marked with an asterisk).
All quoted errors are 1$\sigma$ uncertainties.
}
\label{tab:betcep_magfield}
\begin{tabular}{rcr@{$\pm$}lr@{$\pm$}l}
\hline
\hline
\multicolumn{1}{c}{MJD} &
\multicolumn{1}{c}{Rotation} &
\multicolumn{2}{c}{$\left<B_{\rm z}\right>_{\rm all}$} &
\multicolumn{2}{c}{$\left<B_{\rm z}\right>_{\rm hyd}$} \\
\multicolumn{1}{c}{} &
\multicolumn{1}{c}{Phase} &
\multicolumn{2}{c}{[G]} &
\multicolumn{2}{c}{[G]} \\
\hline
\multicolumn{6}{c}{$\xi^1$\,CMa} \\
\hline
  53475.046 & 0.702 & 282 & 42 & 280 & 44 \\
  53506.971 & 0.351 & 278 & 43 & 330 & 45 \\
  54061.325 & 0.715 & 287 & 42 & 360 & 45 \\
  54107.266 & 0.795 & 312 & 43 & 319 & 46 \\
  54114.028 & 0.898 & 309 & 35 & 347 & 38 \\
  54114.182 & 0.969 & 364 & 35 & 382 & 47 \\
  54116.108 & 0.853 & 307 & 45 & 276 & 58 \\
  54155.086 & 0.738 & 308 & 47 & 349 & 35 \\
  54343.371 & 0.132 & 345 & 11 & 379 & 15 \\
  54345.338 & 0.034 & 366 & 11 & 400 & 12 \\
  54345.414 & 0.069 & 340 & 11 & 378 & 18 \\
  54548.982 & 0.476 & 277 & 55 & 297 & 87 \\
  54549.995 & 0.941 & 380 & 37 & 332 & 55 \\
 $^{\ast}$54722.274 & 0.991 & 386 & 39 & \multicolumn{2}{c}{} \\
  55107.342 & 0.678 & 229 & 30 & 302 & 44 \\
  55109.325 & 0.589 & 206 & 31 & 233 & 52 \\
  55113.224 & 0.378 & 203 & 44 & 320 & 65 \\
  55135.200 & 0.461 & 213 & 39 & 240 & 59 \\
  55150.342 & 0.409 & 176 & 51 & 322 & 76 \\
  55153.340 & 0.784 & 295 & 61 & 470 & 94 \\
  55159.329 & 0.532 & 207 & 29 & 254 & 41 \\
  55163.085 & 0.256 & 282 & 35 & 389 & 56 \\
  55164.092 & 0.718 & 272 & 45 & 416 & 88 \\
  55165.106 & 0.183 & 301 & 39 & 431 & 64 \\
  55168.091 & 0.553 & 232 & 44 & 174 & 59 \\
 $^{\ast}$55201.279 & 0.781 & 297 & 26 & \multicolumn{2}{c}{} \\
\hline
\multicolumn{6}{c}{$\alpha$\,Pyx} \\
\hline
  54082.341 & 0.816 &   142 & 48 &    219 & 60 \\
  54109.150 & 0.200 &   132 & 50 &    184 & 60 \\
  55107.378 & 0.362 &     5 & 31 &  $-$29 & 48 \\
  55118.347 & 0.792 &   120 & 28 &     89 & 34 \\
  55120.351 & 0.418 & $-$14 & 26 &  $-$35 & 44 \\
  55168.167 & 0.371 &    24 & 39 &  $-$26 & 68 \\
  55171.162 & 0.308 &    40 & 45 &     42 & 69 \\
\hline
\multicolumn{6}{c}{$\epsilon$\,Lup} \\
\hline
  54344.998 & & $-$156 & 34 & $-$128 & 36 \\
  55225.268 & & $-$130 & 94 & $-$191 & 129 \\
  55226.272 & &  $-$33 & 39 &  $-$40 & 52 \\
  55227.285 & & $-$104 & 44 & $-$185 & 69 \\
  55228.324 & &   $-$7 & 42 &      7 & 51 \\
  55258.206 & & $-$147 & 46 & $-$162 & 54 \\
  55259.260 & & $-$105 & 41 & $-$158 & 66 \\
\hline
\multicolumn{6}{c}{15\,CMa} \\
\hline
  54107.318 & 0.085 &   163 & 52 &    157 &  58 \\
  54345.372 & 0.917 &   149 & 19 &    123 &  27 \\
  55107.326 & 0.192 &    86 & 26 &    113 &  55 \\
  55109.371 & 0.354 & $-$41 & 36 & $-$130 &  52 \\
  55112.380 & 0.592 & $-$92 & 48 &  $-$88 &  74 \\
  55150.330 & 0.594 & $-$75 & 54 &  $-$49 &  78 \\
  55159.295 & 0.304 &    35 & 34 &     13 &  51 \\
  55163.101 & 0.605 & $-$25 & 39 &  $-$45 &  58 \\
  55164.104 & 0.684 &    31 & 49 &  $-$24 &  61 \\
  55165.118 & 0.764 &    76 & 41 &     19 &  60 \\
  55168.104 & 0.000 &   128 & 42 &    126 &  65 \\
  55170.090 & 0.158 &   138 & 52 &     90 & 120 \\
  55171.133 & 0.240 &    21 & 42 &  $-$12 &  67 \\
  55173.143 & 0.399 & $-$58 & 88 &  $-$64 & 122 \\
  55177.323 & 0.730 &     7 & 33 &      0 &  62 \\
\hline
\end{tabular}
\end{table}

\begin{table}\small
\centering
\caption{
%Magnetic field measurements of $\beta$\,Cephei and SPB stars with FORS\,2.
%Phases are calculated according to the ephemeris of 
%Stahl et al.\ (\cite{stahl:1996}), ${\rm JD} = 2\,448\,833.0 + 15.422~{\rm E}$.
Magnetic field measurements of SPB stars with FORS\,1/2.
All quoted errors are 1$\sigma$ uncertainties.
}
\label{tab:spb_magfield}
\begin{tabular}{ccr@{$\pm$}lr@{$\pm$}l}
\hline
\hline
\multicolumn{1}{c}{MJD} &
\multicolumn{1}{c}{Rotation} &
\multicolumn{2}{c}{$\left<B_{\rm z}\right>_{\rm all}$} &
\multicolumn{2}{c}{$\left<B_{\rm z}\right>_{\rm hyd}$} \\
\multicolumn{1}{c}{} &
\multicolumn{1}{c}{Phase} &
\multicolumn{2}{c}{[G]} &
\multicolumn{2}{c}{[G]} \\
\hline
\multicolumn{6}{c}{33\,Eri} \\
\hline
  52971.071 & 0.599 & $-$122 & 64 & $-$120 & 68 \\
  53574.415 & 0.157 &  $-$14 & 33 &  $-$16 & 36 \\
  53630.250 & 0.796 &  $-$34 & 27 &  $-$32 & 29 \\
  54086.175 & 0.135 &  $-$30 & 54 &  $-$45 & 61 \\
  54343.301 & 0.098 &     67 & 60 &     83 & 65 \\
  55107.154 & 0.105 &    116 & 39 &    166 & 49 \\
  55108.197 & 0.929 &  $-$25 & 43 &  $-$43 & 56 \\
  55109.350 & 0.822 &      5 & 35 &     10 & 49 \\
  55110.181 & 0.471 &  $-$98 & 48 & $-$138 & 70 \\
  55111.222 & 0.284 &  $-$39 & 45 &  $-$52 & 69 \\
  55112.347 & 0.164 &      0 & 43 &  $-$20 & 72 \\
  55113.188 & 0.821 &     25 & 44 &     62 & 73 \\
  55120.099 & 0.223 &  $-$63 & 55 & $-$177 & 79 \\
  55135.184 & 0.013 &     90 & 52 &    112 & 64 \\
  55149.176 & 0.948 &    117 & 37 &    119 & 44 \\
  55150.302 & 0.829 &     12 & 46 &   $-$7 & 72 \\
  55161.089 & 0.259 & $-$126 & 46 & $-$168 & 55 \\
  55163.031 & 0.777 &  $-$41 & 44 &  $-$18 & 59 \\
\hline
\multicolumn{6}{c}{HY\,Vel} \\
\hline
  53002.141 & & $-$180 & 57 & $-$199 & 61 \\
  53143.986 & &  $-$48 & 60 &  $-$53 & 66 \\
  54108.348 & & $-$198 & 55 & $-$191 & 58 \\
  55107.361 & &    34 & 59 &  $-$61 & 79 \\
  55112.367 & & $-$25 & 51 &  $-$58 & 69 \\
  55118.320 & &   133 & 62 &    162 & 83 \\
  55120.324 & &    63 & 33 &     59 & 48 \\
  55168.150 & & $-$99 & 38 & $-$147 & 64 \\
  55169.349 & &    47 & 41 &     78 & 73 \\
  55171.150 & &    40 & 59 &  $-$11 & 88 \\
  55173.269 & &    17 & 51 &  $-$19 & 68 \\
  55177.341 & &   117 & 35 &    160 & 49 \\
  55181.141 & &    42 & 36 &     35 & 54 \\
  55182.280 & &    67 & 38 &    144 & 53 \\
  55189.176 & &    19 & 44 &     30 & 68 \\
\hline
\end{tabular}
\end{table}

A frequency analysis  was performed  on the longitudinal magnetic field 
measurements $\left<B_{\rm z}\right>_{\rm all}$ (which generally show smaller sigmas)
available from our previous \citep{Hubrig2006,Hubrig2009} and 
the current studies using a non-linear least-squares fit of the 
multiple harmonics utilizing the Levenberg-Marquardt 
method \citep{Press1992} with an optional possibility of pre-whitening the trial harmonics.  To detect 
the most probable period, we calculated the frequency spectrum for the same harmonic with a number of trial 
frequencies by solving the linear least-squares problem. At each trial frequency we performed a statistical 
test of the null hypothesis for the absence of periodicity \citep{Seber77}, i.e.\ testing that all harmonic 
amplitudes are at zero. The resulting F-statistics can be thought of as the total sum including covariances of the 
ratio of harmonic amplitudes to their standard deviations, i.e.\ as a signal-to-noise ratio 
\citep{Ilyin2000}.
%Assuming that the noise of the data is normally distributed,
The F-statistics allows to derive 
the false alarm probability of the trial period based on the F-test \citep{Press1992}.  
%A number of similar approaches 
%to the revision of least-squares periodograms has been devised in the literature 
%e.g.\ Cumming (\cite{Cumming2004}), Baluev (\cite{Baluev2008}), and Zechmeister \& K\"urster (\cite{ZechmeisterKuerster2009}). 
Periodicity was found for four out of the studied six stars. 
%%the resulting amplitude spectra displayed dominant peaks with 
%%equivalent periods of: 2.18\,d 
%%for the $\beta$\,Cephei star $\xi^1$\,CMa,
%%a period of 12.64\,d for the $\beta$\,Cephei star 15\,CMa,
%%a period of 3.20\,d for the $\beta$\,Cephei star $\alpha$\,Pyx,
%%and a period of 1.28\,d for the SPB star 33\,Eri. For two stars in our sample, $\epsilon$\,Lup
%%and HY\,Vel, no periodicity was found.
%only for four stars: for $\beta$\,Cephei stars $\xi^1$\,CMa, 15\,CMa and $\alpha$\,Pyx, and for the SPB star 33\,Eri.
The derived ephemeris for the detected periods are 

\begin{eqnarray}
\xi^1\,{\rm CMa}: \left<V\&I\right>^{\rm max} &= {\rm MJD}55140.73332 \pm 0.03794 + 2.17937 \pm 0.00012 E  \nonumber \\
15\,{\rm CMa}: \left<V\&I\right>^{\rm max} &= {\rm MJD}55168.09911 \pm 0.16667 + 12.64115 \pm 0.00822 E  \nonumber \\
\alpha\,{\rm Pyx}: \left<V\&I\right>^{\rm max} &= {\rm MJD}55144.59481 \pm 0.04105 + 3.19779 \pm 0.00019 E  \nonumber \\
33\,{\rm Eri}: \left<V\&I\right>^{\rm max} &= {\rm MJD}55123.65285 \pm 0.03243 + 1.27947 \pm 0.00005 E  \nonumber
%\xi^1~{\rm CMa}:& \nonumber \\
 %\left<V\&I\right>^{\rm max} =&{\rm MJD}55140.733 \pm 0.038 +  2.17937 \pm 0.00012 E  \nonumber \\
%15~{\rm CMa}:& \nonumber \\
 %\left<V\&I\right>^{\rm max} =&{\rm MJD}55168.099 \pm 0.167 + 12.64115 \pm 0.00822 E  \nonumber \\
%\alpha~{\rm Pyx}:& \nonumber \\
 %\left<V\&I\right>^{\rm max} =&{\rm MJD}55144.595 \pm 0.041 + 3.19779 \pm 0.00019 E  \nonumber \\
%33~{\rm Eri}:& \nonumber \\
 %\left<V\&I\right>^{\rm max} =&{\rm MJD}55123.653 \pm 0.032 + 1.27947 \pm 0.00005 E  \nonumber 
\end{eqnarray}

%Captions:
%The frequency spectrum (with a single harmonic of the first degree) of the data with the highest peak at frequency f=0.13745 or P=7.27559 days with the window function overplotted in the red.

\begin{figure*}
\centering
\includegraphics[width=1.00\textwidth]{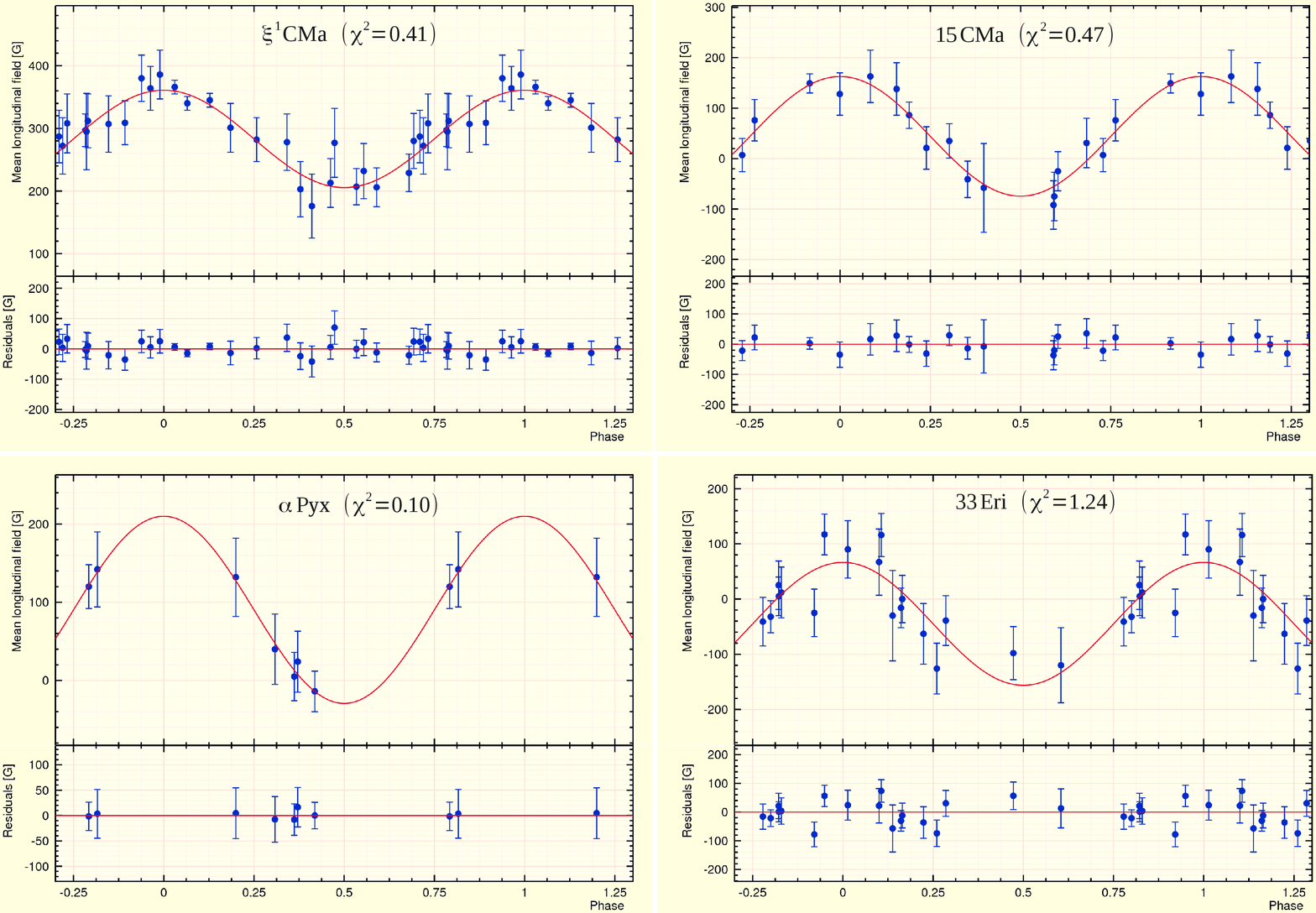}
%\includegraphics[width=1.00\textwidth]{fit_2prev.eps}
%\includegraphics[width=0.45\textwidth]{15cma.pc.ps}
%\vskip -0.5mm
%\includegraphics[width=0.45\textwidth]{HD46328.pr.ps}
%\includegraphics[width=0.45\textwidth]{15cma.pr.ps}
%\includegraphics[width=0.45\textwidth]{HD74575.pc.ps}
%\includegraphics[width=0.45\textwidth]{33eri.pc.ps}
%\vskip -0.5mm
%\includegraphics[width=0.45\textwidth]{HD74575.pr.ps}
%\includegraphics[width=0.45\textwidth]{33eri.pr.ps}
\caption{
Phase diagrams with the best sinusoidal fit for the longitudinal magnetic field measurements.
The residuals (Observed -- Calculated) are shown in the lower panels. The deviations are mostly of the same 
order as the error bars, and no systematic trends are obvious, which justifies a single sinusoid as a 
fit function.
%The fits correspond to $\xi^1$\,CMa (upper left), 
%15\,CMa (upper right),
%$\alpha$\,Pyx (lower left),
%and 33\,Eri (lower right).
}
\label{fig:all}
\end{figure*}

The logbook of the new FORS\,2 and the old, revisited, FORS\,1 
spectropolarimetric observations is presented
in Tables~\ref{tab:betcep_magfield} and \ref{tab:spb_magfield}.
%in Tables~\ref{tab:xi1cma_magfield} to \ref{tab:hyvel_magfield}.
In the first column we indicate the MJD value at mid exposure. The phases of the measurements of 
the magnetic field, if available,
are listed in Column~2. In Columns~3 and 4 we present the longitudinal magnetic 
field $\left<B_{\rm z}\right>_{\rm all}$ measured using the whole spectrum and the longitudinal magnetic field 
$\left<B_{\rm z}\right>_{\rm hyd}$ using only the hydrogen lines. 
Phase diagrams of the data folded with the determined periods are presented in Fig.~\ref{fig:all}.
%Figs.~\ref{fig:1}-~\ref{fig:4}.
The quality of our fits is described by a reduced $\chi^2$-value which appears in the four panels
of Fig.~\ref{fig:all}.
%which is 0.41 for $\xi^1$\,CMa, 0.47 for 
%15\,CMa, 0.10 for $\alpha$\,Pyx, and 1.24 for 33\,Eri.

%The achieved signal-to-noise ratios in the wavelength region 5000-5500\,\AA{} were 550 for HD\,108 and 360 for HD\,191612(???).

\section{Characterisation of the magnetic field geometry for the stars with determined periods}

\begin{table}
\centering
\caption{
Magnetic field models for the stars with detected periods.
}
\label{tab:dipvals}
\begin{tabular}{cc|r@{$\pm$}lr@{$\pm$}lr@{$\pm$}lr@{$\pm$}l}
\hline
\hline
\multicolumn{2}{c|}{Object} &
\multicolumn{2}{c}{$\xi^1$\,CMa} & 
\multicolumn{2}{c}{15\,CMa} & 
\multicolumn{2}{c}{$\alpha$\,Pyx} & 
\multicolumn{2}{c}{33\,Eri} \\
\hline
$\overline{\left< B_{\rm z}\right>}$ & [G]           & 281.6 &  4.3 & 45.3 &  7.1 & 90.3 &  5.1 & $-$44.1 & 15.3 \\ 
$A_{\left< B_{\rm z}\right>}$        & [G]           & 80.3 &  5.8 & 116.3 & 12.1 & 119.7 & 12.8 & 112.0 & 25.5 \\
$v$\,sin\,$i$                        & [km s$^{-1}$] & 9 & 2 & 34 & 4 & 11 & 2 & 25 & 4 \\
$R$                                  & [R$_{\odot}$] & 7.1 & 0.9 & 10.0 & 1.5 & 6.3 & 1.0 & 2.5 & 0.3 \\
$v_{\rm eq}$                         & [km s$^{-1}$] & 165 & 21 & 40 &  6 & 100 & 16 & 99 & 12 \\
$i$                                  & [$^{\circ}$]  & 3.1 &  0.8 & 58.1 & 17.6 & 6.3 &  1.5 & 14.6 &  3.0 \\
$\beta$                              & [$^{\circ}$]  & 79.1 &  2.8 & 57.9 & 18.3 & 85.2 &  1.3 & 84.1 &  2.7 \\
%B$_{\rm d}$                         & [G]           & 5300 & 1100 & 570 &  190 & 3800 &  700 & 1500 &  400 \\
B$_{\rm d}$                         & [G]           & 5300 & 1100 & 570 &   50 & 3800 &  500 & 1500 &  350 \\
\hline
\end{tabular}
\end{table}

The determination of the fundamental parameters and the description of
the pulsating properties of all studied stars were presented by
\citeauthor{Hubrig2009} (\citeyear{Hubrig2009}, Tables~1a, 1b, and 4).
The most simple modeling of the magnetic field geometry
is based on the assumption that the studied stars are oblique dipole rotators, i.e.\ their magnetic field 
can be approximated by a dipole with the magnetic axis inclined to the rotation axis. 
%In the following we discuss the magnetic field geometry for stars with detected periodicity.

The magnetic dipole axis tilt $\beta$ is constrained by 

\begin{equation} 
r = \frac{\left< B_{\rm z}\right>^{\rm min}}{\left< B_{\rm z}\right>^{\rm max}} 
  = \frac{\cos \beta \cos i - \sin \beta \sin i}{\cos \beta \cos i + \sin \beta \sin i}, 
\end{equation} 
 
\noindent 
so that the obliquity angle $\beta$ is given by
 
\begin{equation} 
\beta =  \arctan \left[ \left( \frac{1-r}{1+r} \right) \cot i \right]. 
\label{eqn:4} 
\end{equation} 

%{\bf In Table~\ref{tab:dipvals} we show for each star with detected periodicity in Cols.~2 and 3
%the mean value $\overline{\left< B_{\rm z}\right>}$ and the amplitude of the field variation $A_{\left< B_{\rm z}\right>}$.
%In Col.~4 we present $v \sin i$ values published recently by \citet{Lefever2010} for $\xi^1$\,CMa and 
%15\,CMa and from \citet{Hubrig2009} for $\alpha$\,Pyx and 33\,Eri. 
%The radius values in Col.~5 were taken from \citet{Hubrig2009}. 
%The radius of 15\,CMa 
%($R$=10.0$\pm$1.5\,R$_\odot$) was  derived in the same way as in \citet{Hubrig2009} by adopting the 
%values of $T_{\rm eff}$ and $\log g$ from \citet{Lefever2010}.
%In the last four columns we list the $v_{\rm eq}$ and the parameters of the magnetic field dipole models.
%The polar field strength $B_{\rm d}$ in the last column was calculated following 
%\citet{Preston1969} using limb darkening parameters from \citet{DiazCordoves1995}.}

In Table~\ref{tab:dipvals} we show for each star with detected periodicity in Rows 2 and 3
the mean value $\overline{\left< B_{\rm z}\right>}$ and the amplitude of the field variation $A_{\left< B_{\rm z}\right>}$.
In Row~4 we present $v \sin i$ values published recently by \citet{Lefever2010} for $\xi^1$\,CMa and 
15\,CMa and from \citet{Hubrig2009} for $\alpha$\,Pyx and 33\,Eri. 
The radius values in Row~5 were taken from \citet{Hubrig2009}. 
The radius of 15\,CMa 
($R$=10.0$\pm$1.5\,R$_\odot$) was  derived in the same way as in \citet{Hubrig2009} by adopting the 
values of $T_{\rm eff}$ and $\log g$ from \citet{Lefever2010}.
In the last four rows we list the $v_{\rm eq}$ and the parameters of the magnetic field dipole models.
The polar field strength $B_{\rm d}$ in the last row was calculated following 
\citet{Preston1969} using limb darkening parameters from \citet{DiazCordoves1995}.

\section{Discussion}

%In our first publication on the magnetic survey of pulsating B-type stars (Hubrig et al.\ \cite{Hubrig2006}), 
%we announced detections of weak mean longitudinal 
%magnetic fields of the order of a few hundred~Gauss in 13~SPB stars and in 
%the $\beta$~Cephei star $\xi^1$~CMa. Among the very few magnetic $\beta$\,Cephei stars, 
%$\xi^1$\,CMa showed the largest mean longitudinal field of the order of 300\,G. 
Using FORS\,1/2 and SOFIN longitudinal magnetic field
measurements collected in our recent studies, we were able to determine rotation
periods and constrain the field geometry of two $\beta$\,Cephei stars,
one candidate $\beta$\,Cephei star, and one SPB star.
The dipole model provides a satisfactory fit to the data and
among the very few presently known magnetic $\beta$\,Cephei stars, 
$\xi^1$\,CMa and $\alpha$\,Pyx
possess the largest magnetic fields, with a dipole strength of several kG. \citet{Briquet2007} 
discussed the position of SPB and chemically peculiar Bp stars in the H-R diagram. indicating that 
the group of Bp stars is significantly younger than the group of SPB stars. A similar conclusion was 
deduced by \citet{Hubrig2007}, who studied the evolution of magnetic fields in Ap and Bp stars with
definitely determined magnetic field geometries across the main sequence. 
The vast majority of Bp stars exhibits a smooth, single-wave variation of 
the longitudinal magnetic field during the stellar rotation cycle. These observations are considered as evidence 
for a dominant dipolar contribution to the magnetic field topology. It is of interest that 
the study of \citet{Hubrig2007} indicates the prevalence of larger obliquities $\beta$, namely $\beta$$>$60$^{\circ}$, 
in more massive stars.
The magnetic field models for the three $\beta$\,Cephei stars and the one SPB star presented in this 
work confirm this trend.

The insufficient knowledge of the strength, geometry, and time variability of magnetic fields in hot 
pulsating stars prevented until now important theoretical studies on the impact of magnetic fields on stellar rotation, 
pulsations, and element diffusion. Although it is expected that the magnetic field can distort the frequency patterns
(e.g.\ \citealt{Hasan2005}),
such a perturbation is not yet detected in hot pulsating stars.
Splitting of non-radial pulsation modes was observed for 15\,CMa \citep{Shobbrook2006},  but
the identification of these modes  is  still pending.
The magnetic $\beta$\,Cephei star sample indicates that 
they all share common properties: 
%(Hubrig et al.\ 2009, AN, 330, 317) 
they are N-rich targets (e.g.,
%Morel et al.\ 2006, A\&A, 457, 651; 
\citealt{Morel2008}) and, as discussed by \citet{Hubrig2009},
their pulsations are dominated 
by a non-linear dominant radial mode (see also \citealt{Saesen2006} for $\xi^1$\,CMa). The 
presence of a magnetic field might consequently play an important role to explain such a distinct 
behaviour of magnetic $\beta$\,Cephei stars. More precisely, chemical abundance anomalies are commonly believed to be 
due to radiatively-driven microscopic diffusion in stars rotating sufficiently slowly to allow such a 
process to be effective. However, we need an additional clue to account for the fact that both normal 
and nitrogen-enriched slowly rotating stars are observed. 
%(Morel et al.\ 2008, A\&A, 481, 453). 
The presence of a magnetic field is a very plausible explanation, as it can add to the stability of the 
atmosphere, allowing diffusion processes to occur \citep{Michaud1970}.
On the other hand, 
among the studied stars, apart from the star 15\,CMa with rather low $v_{\rm eq} = 40\pm 6$\,km\,s$^{-1}$, 
the other three magnetic pulsating stars 
rotate much faster up to $v_{\rm eq} = 165\pm 21$\,km\,s$^{-1}$ for $\xi^1$\,CMa with the 
strongest
magnetic field, indicating that these stars are not truly slowly rotating stars, but seen close to pole-on.
%with $v_{\rm eq} = 99\pm 12$\,km\,s$^{-1}$ for 33\,Eri, $v_{\rm eq} = 100\pm 16$\,km\,s$^{-1}$
%for $\alpha$\,Pyx, and the highest velocity $v_{\rm eq} = 165\pm 21$\,km\,s$^{-1}$ for $\xi^1$\,CMa with the 
%strongest
%magnetic field, indicating that these stars are not truly slowly rotating stars, but seen close to pole-on.
Obviously, the topic of mixing signatures is not understood theoretically yet and more computational work as well as 
future additional observational validation of our results are needed to understand the link between 
the presence of a magnetic field, rotation, pulsating characteristics, and abundance peculiarities.  

\acknowledgments

T.M. acknowledges financial 
support from Belspo for contract PRODEX GAIA-DPAC.


\begin{thebibliography}{}

%\bibitem[Catala et al.(2007)]{Catala2007}
%Catala, C., et al.\ 2007,
%\aap, 462, 293 

%\bibitem[Baluev(2008)]{Baluev2008}
%Baluev, R. V.\ 2008, MNRAS 385, 1279

%\bibitem[Cumming(2004)]{Cumming2004}
%Cumming, A.\ 2004, MNRAS 354, 1165

\bibitem[Babcock(1958)]{Babcock1958}
Babcock, H.~W., 1958,
\apjs, 3, 141

\bibitem[Briquet et al.(2007)]{Briquet2007}
Briquet, M., Hubrig, S., De Cat, P., Aerts, C., North, P., \& Sch\"oller, M., 2007,
\aap, 466, 269

\bibitem[Diaz-Cordov\'es et al.(1995)]{DiazCordoves1995}
Diaz-Cordov\'es, J., Claret, A., \& Gimenez, A.\ 1995,
A\&AS, 110, 329

\bibitem[Hasan et al.(2005)]{Hasan2005}
Hasan, S.~S., Zahn, J.-P., \& Christensen-Dalsgaard, J.\ 2005,
\aap, 444, L29

\bibitem[Henrichs et al.(2000)]{Henrichs2000}
Henrichs, H.~F., Neiner, C., Hubert, A.~M., Floquet, M., \& the MuSiCoS Team 2000,
in ASP Conf.\ Ser.\ Vol.\ 214, The Be Phenomenon in Early-Type Stars,
eds.\ M.A.\ Smith \& H.F.\ Henrichs, 372

\bibitem[Hubrig et al.(2004a)]{Hubrig2004a}
Hubrig, S., Kurtz, D.~W., Bagnulo, S., Szeifert, T., Sch{\"o}ller, M., Mathys, G., \& Dziembowski, W.~A.\ 2004a,
A\&A, 415, 661

\bibitem[Hubrig et al.(2004b)]{Hubrig2004b} 
Hubrig, S., Szeifert, T., Sch\"oller, M., Mathys, G., \& Kurtz, D.~W.\ 2004b,
A\&A, 415, 685

\bibitem[Hubrig et al.(2006)]{Hubrig2006}
Hubrig, S., Briquet, M., Sch{\"o}ller, M., De Cat, P., Mathys, G., \& Aerts, C.\ 2006,
MNRAS, 369, L61

\bibitem[Hubrig et al.(2007)]{Hubrig2007}
Hubrig, S., North, P., \& Sch\"oller, M., 2007,
AN 328, 475

\bibitem[Hubrig et al.(2009)]{Hubrig2009}
Hubrig, S., Briquet, M., De Cat, P., Sch{\"o}ller, M., Morel, T., \& Ilyin, I.\ 2009,
AN, 330, 317

\bibitem[Ilyin(2000)]{Ilyin2000}
Ilyin, I.~V.\ 2000,
Numerical methods for the data analysis, Manuscript, 132 pages

\bibitem[Lefever et al.(2010)]{Lefever2010}
Lefever, K., Puls, J., Morel, T., Aerts, C., Decin, L., \& Briquet, M.\ 2010,
\aap, 515, A74

\bibitem[Michaud(1970)]{Michaud1970}
Michaud, G.\ 1970,
\apj, 160, 641

\bibitem[Morel et al.(2008)]{Morel2008}
Morel, T., Hubrig, S., \& Briquet, M.\ 2008,
A\&A, 481, 453

\bibitem[Neiner et al.(2003a)]{Neiner2003a}
Neiner, C., Geers, V.C., Henrichs, H.F., Floquet, M., Fr{\'e}mat, Y., Hubert, A.-M., Preuss, O., \& Wiersema, K.\ 2003a,
A\&A, 406, 1019

\bibitem[Neiner et al.(2003b)]{Neiner2003b}
%Neiner, C., Henrichs, H.F., Floquet, M., et al.\ 2003b,
Neiner, C., et al.\ 2003b,
A\&A, 411, 565

\bibitem[Press et al.(1992)]{Press1992}
Press, W.~H., Teukolsky, S.~A., Vetterling, W.~T., \& Flannery, B.~P.\ 1992,
Numerical Recipes, 2nd ed.\ (Cambridge University Press: Cambridge)

%\bibitem[Preston(1967)]{Preston1967} 
%Preston, G.~W.\ 1967,
%ApJ, 150, 547

\bibitem[Preston(1969)]{Preston1969} 
Preston, G.~W.\ 1969,
ApJ, 156, 967

%\bibitem[Przybilla \& Nieva(2010)]{PrzybillaNieva2010}
%Przybilla, N., \& Nieva, M.-F.\ 2010,
%in IAU Symp.\ 272, Active OB stars -- structure, evolution, mass-loss, and critical limits,
%eds.\ C.\ Neiner, G.\ Wade, G.\ Meynet, \& G.\ Peter, {\sl in press}

\bibitem[Rudy \& Kemp(1978)]{RudyKemp1978}
Rudy, R.~J., \& Kemp, J.~C., 1978,
\mnras, 183, 595

\bibitem[Saesen(2006)]{Saesen2006}
Saesen, S., Briquet, M., \& Aerts, C.\ 2006,
Comm.\ in Asteroseismology, 147, 109

\bibitem[Seber(1977)]{Seber77}
Seber, G.~A.~F.\ 1977,
Linear Regression Analysis (Wiley: New York) 

\bibitem[Shibahashi \& Aerts(2000)]{ShibahashiAerts2000}
Shibahashi, H., \& Aerts, C., 2000,
\apjl, 531, L143

\bibitem[Shobbrook et al.(2006)]{Shobbrook2006}
Shobbrook, R.~R., Handler, G., Lorenz, D., \& Mogorosi, D.\ 2006,
\mnras, 369, 171

\bibitem[Stankov \& Handler(2005)]{StankovHandler2005}
Stankov, A., \& Handler, G.\ 2005,
\apjs, 158, 193

\bibitem[Tuominen et al.(1999)]{Tuominen1999}
Tuominen, I., Ilyin, I., \& Petrov, P.\ 1999,
in Astrophysics with the NOT, eds.\ H.\ Karttunen \& V.\ Piirola, University of Turku, Tuorla Observatory, 47

%\bibitem[Zechmeister \& K\"urster(2009)]{ZechmeisterKuerster2009}
%Zechmeister, M., K\"urster, M.\ 2009, A\&A 496, 577

\end{thebibliography}
\end{document}